\documentclass[preprintnumbers,amsmath,amssymb,sublabel,prl,twocolumn,showpacs]{revtex4}
\usepackage{amsfonts}
\usepackage{amssymb}
\usepackage{amsmath}
\usepackage{amsthm}
\usepackage{mathrsfs}
\usepackage{bm}
\usepackage{graphicx}
\usepackage{dcolumn}
\usepackage{multirow}
\usepackage{float}
\usepackage{sublabel}
\usepackage{times,epsfig}

\begin{document}
\title{Minimal sets determining universal and phase-covariant quantum cloning}

\author{Li Jing$^1$, Yi-Nan Wang$^1$, Han-Duo Shi$^1$, Liang-Zhu Mu$^1$\footnote{muliangzhu@pku.edu.cn}
and Heng Fan$^2$\footnote{hfan@iphy.ac.cn}}
\affiliation{%
$^1$School of Physics, Peking University, Beijing 100871, China\\
$^2$Beijing National Laboratory for Condensed Matter Physics,
Institute of Physics, Chinese Academy of Sciences, Beijing
100190, China
}%

\date{\today}

\begin{abstract}
We study the minimal input sets which can determine completely the universal and the
phase-covariant quantum cloning machines. We find that the universal quantum cloning machine,
which can copy arbitrary input qubit equally well, however
can be determined completely by only four input states located at the four vertices of a tetrahedron.
The phase-covariant quantum cloning machine, which can copy all qubits located
on the equator of the Bloch sphere, can be determined by three equatorial qubits
with equal angular distance. These results sharpen further the well-known results
that BB84 states and six-states used in quantum cryptography can determine completely the phase-covariant and universal
quantum cloning machines. This concludes the study of the power of universal and phase-covariant
quantum cloning, i.e., from minimal input sets necessarily to full input sets by definition.
This can simplify dramatically the testing of whether the quantum clone machines are successful or not,
we only need to check that the minimal input sets can be cloned optimally.
\end{abstract}

\pacs{03.67.Ac, 03.67.Dd, 03.65.Aa}


\maketitle

\emph{Introduction.}---
No-cloning theorem, which states that an unknown quantum state cannot be cloned perfectly \cite{nocloning},
is fundamental for quantum information science \cite{Nielsen-Chuang2000}.
However, one can attempt to clone quantum states imperfectly.
In the past years, different schemes of quantum cloning are proposed,
and various quantum cloning machines are designed for different tasks[2-5].
The quantum cloning machine was first proposed to clone an arbitrary qubit
to two equally well qubits \cite{UQCM}, both of them are not identical
but close to the input qubit. The quality of the quantum cloning does
not depend on the input qubit, so it is named as universal quantum cloning machine (UQCM).
This cloning machine is proved to be optimal in the sense
that the fidelity between the input qubit and one of
the two output qubits is optimal\cite{BDEF}.
The UQCM is extended to, such as the higher
dimensional case \cite{d_dim}, the case with
$M$ identical input states to $N$ equally copies \cite{GM1997},
and some more general cases \cite{review,CerfSecurity,Fan,BCDM,TwoBases,FGGNP,1-Nphase,FIMW,Bechmann}
including the recent proposed unified forms \cite{Wangunified,Xiong}.

\begin{figure}[!h]
\includegraphics[width=4cm]{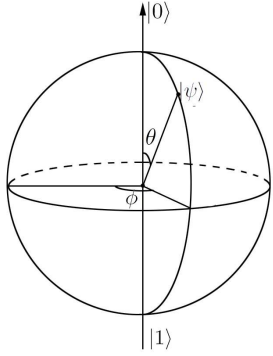}
\caption{(color online). A qubit in Bloch sphere $|\psi \rangle
=\cos ({\theta /2})|0\rangle +\sin ({\theta /2})e^{i\phi }|1\rangle $ which is characterized
by amplitude parameter $\theta $ and phase parameter $\phi $. A equatorial qubit
is the qubit located on the equator of the Bloch sphere. }
\label{qubit}
\end{figure}

A qubit can be represented as, $|\psi \rangle =\cos ({\theta /2})|0\rangle +\sin ({\theta /2})e^{i\phi }|1\rangle $,
where $\theta \in [0,\pi ], \phi \in [0, 2\pi\}$, it corresponds to a point in the Bloch sphere, see FIG.\ref{qubit}.
For the UQCM, the input can be arbitrary qubits,
the fidelity is optimal and does not depend on the input qubit. However,
if we restrict the input state to the equatorial qubit which
is located in the equator of the Bloch sphere $|\psi \rangle =(|0\rangle +e^{i\phi }|1\rangle )/\sqrt {2}$,
one can find that we can clone it better by a different
quantum cloning machine. This cloning machine is phase-covariant in the sense that
the quality of the cloning, similarly quantified by the fidelity, does not depend on the phase parameter $\phi $.
This is the phase-covariant quantum cloning machine (PQCM)\cite{BCDM,1-Nphase,FIMW}.

One important application of quantum cloning machines is to analyze the security of some protocols of quantum key distribution (QKD).
The reason is that a simple quantum attack for eavesdropper is to keep one copy of the quantum state encoding secret key
while still send another copy to the legitimate receiver. For the well-known BB84 protocol \cite{BB84},
we use two sets of orthogonal qubits, $\{ |0\rangle ,|1\rangle \},\{ (|0\rangle +|1\rangle )/\sqrt{2},
(|0\rangle -|1\rangle )/\sqrt{2}\}$,
to encode binary secrete key $0$ or $1$. It seems straightforward that BB84 states correspond to
four orthotropic equatorial qubits: $\{ (|0\rangle \pm |1\rangle )/\sqrt{2}\},
\{(|0\rangle \pm i|1\rangle )/\sqrt{2}\} $. Thus at least, we should use
PQCM instead of UQCM for eavesdropping. The point is that it is possible that we can do better.
Surprisingly, it is shown that PQCM is already the optimal one in copying those four orthotropic
equatorial qubits \cite{BCDM}. Similar phenomenon happens in case of six-state QKD \cite{Bruss}, where the
involved six states are BB84 states plus two equatorial qubits $\{(|0\rangle \pm i|1\rangle )/\sqrt{2}\} $.
We can not do better in cloning the six-state QKD than a UQCM which can clone optimally an arbitrary qubit.

This seems not the end. With continuous progress of quantum cloning theoretically and experimentally
\cite{naturephotonics09,Recent1,Recent2,Recent3,Recent4,
Recent5,Recent6,Recent7}, for years, a question is never asked whether BB84 states and six states are
the minimal input sets necessarily for PQCM and UQCM.
In this Letter, we find that they are not! The
minimal input sets which can determine completely the PQCM and UQCM are found.
The minimal input sets contain only three and four states, respectively.
This will conclude the study of input states for PQCM and UQCM,
the lower bounds are identified here while the upper bounds are from their definitions.

The importance of this result is that, experimentally if we find that the quantum cloning machines
can copy the corresponding minimal input sets optimally, we know that they are able to
clone optimally all equatorial qubits and arbitrary qubits, respectively. This simplifies dramatically
the testing step.
Another importance of this result is that from Heisenberg uncertainty principle in quantum mechanics,
similarly from no-cloning theorem, an unknown quantum state with single copy
cannot be completely identified. So it can be expected
that the minimal input sets for cloning machines would have the
same uncertainty as the full input sets from the definitions
of quantum cloning. Thus our results may shed light on both
the fundamental questions of uncertainty principle, state- and phase- estimations \cite{state,phase-est}
and potentially may induce new applications in quantum cryptography
which relies on no-cloning.

\emph{Optimal quantum cloning machine for three-state.}---We first study the case of
PQCM. It is known that PQCM is necessary for input set with only four equatorial states
equivalent to BB84 states. Here, we try
to find whether it is possible to sharpen it further to three states.
So now, the question is whether we can find a set of three equatorial qubits,
the cloning of of these three states cannot be better than a PQCM. On the other hand,
it is simple to find that two equatorial qubits can always be cloned better than
a PQCM does, so the set of three states will be the minimal input set which can
determine the PQCM. It is known that
the optimal fidelity of PQCM is \cite{BCDM,1-Nphase},
\begin{eqnarray}
F_{p}=\frac {1}{2}+\frac {\sqrt {2}}{4}.
\end{eqnarray}
Thus our goal is to find a set of three states, the fidelity of their cloning
is upper bounded by $F_{p}$. It is apparent that this bound is achievable.

A quantum cloning machine generally needs ancillary states, if no ancillary states are
available, it is the economic quantum cloning. In this Letter, we start from the economical cloning
for simplicity. We will then show that ancillary states
will not help to increase the fidelity.

We consider three equatorial qubits represented as,
\begin{eqnarray}
|\psi_i\rangle=(|0\rangle+\mathrm{e}^{\mathrm{i}\phi_i}|1\rangle)/\sqrt{2},
\end{eqnarray}
where $i=1,2,3$ represents three different phases.
The economic quantum cloning transformation is a unitary transformation $U$ on the
input qubit and an initially blank state which the copied qubit will be set in.
Its general form is:
\begin{eqnarray}
&&U|00\rangle=a|00\rangle+b|01\rangle+c|10\rangle+d|11\rangle\nonumber\\
&&U|10\rangle=e|00\rangle+f|01\rangle+g|10\rangle+h|11\rangle,
\label{cloning1}
\end{eqnarray}
where $a,...,h$ are complex parameters to be determined,
which should satisfy the constraints,
\begin{eqnarray}
&&a^*e+b^*f+c^*g+d^*h=0\nonumber\\
&&|a|^2+|b|^2+|c|^2+|d|^2=1\nonumber\\
&&|e|^2+|f|^2+|g|^2+|h|^2=1.
\end{eqnarray}
The first equation shows the orthogonality of the unitary transformation,
the next two equations are the normalization conditions. Now consider
the input state $|\psi \rangle $, by performing the transformation $U$ on $|\psi0\rangle$,
we obtain the density matrix for the whole system constituted by qubits $A$ and $B$,
$\rho _{AB}=U|\psi0\rangle\langle\psi0|U^\dagger $.
Then we can trace out one of the particles to get one-particle reduced density matrices,
$\rho_A$ or $\rho_B$, they are two copies from the original input state $|\psi \rangle $.
To quantify the quality of the cloning machine,
we use the fidelities, $F_A(\phi)=\langle\psi|\rho_A|\psi\rangle$ and $F_B(\phi)=\langle\psi|\rho_B|\psi\rangle$, to evaluate
the the distance between the input and two copies.
As for our cloning machine (\ref{cloning1}), they are in form,
\begin{eqnarray}
F_A(\phi)=&&\frac{1}{2}+\frac{1}{2}\mathrm{Re}[ac^*\mathrm{e}^{\mathrm{i}\phi}+ag^*+ec^*\mathrm{e}^{2\mathrm{i}\phi}+eg^*\mathrm{e}^{\mathrm{i}\phi}\nonumber\\
&&+bd^*\mathrm{e}^{\mathrm{i}\phi}+fh^*\mathrm{e}^{\mathrm{i}\phi}+fd^*\mathrm{e}^{2\mathrm{i}\phi}+bh^*].
\end{eqnarray}
The expression of $F_B$ is obtained just by interchanging $b\leftrightarrow c$ and $f\leftrightarrow g$.

As in term of phase-covariant,
we let the fidelities do not depend on input states. So for these three equatorial qubits, we should have,
\begin{align}
F_A(\phi_1)&=F_A(\phi_2)=F_A(\phi_3)\nonumber\\
F_B(\phi_1)&=F_B(\phi_2)=F_B(\phi_3).
\end{align}
So we can rewrite the fidelity $F_A$ in the form,
\begin{eqnarray}
F_A=\lambda_1\cos(2\phi+\psi_1)+\lambda_2\cos(\phi+\psi_2)+\lambda_3,
\end{eqnarray}
where $\lambda_i$ are independent real numbers. The explicit expressions of these parameters are:
$\lambda_1=\frac{1}{2}|ec^*+fd^*|$,
$\psi_1=\arg(ec^*+fd^*)$,
$\lambda_2=\frac{1}{2}|ac^*+eg^*+bd^*+fh^*|$,
$\psi_2=\arg(ac^*+eg^*+bd^*+fh^*)$,
$\lambda_3=\frac{1}{2}Re(ag^*+bh^*)+\frac{1}{2}$.

\begin{figure}[!h]
\includegraphics[width=4.5cm]{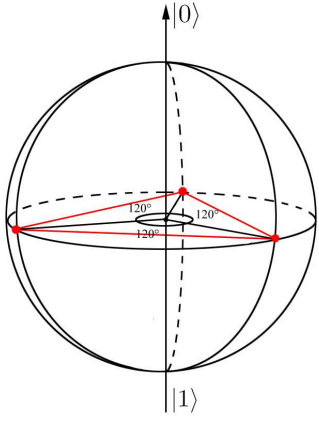}
\caption{(color online). Three equatorial qubits with equal
relative phases $\phi_1=0^\circ,\phi_2=120^\circ$ and $\phi_3=240^\circ $ which
determine a PQCM.}
\label{figure3qubit}
\end{figure}

Then we study three states with 120$^\circ$
intersection angles, that is, $\phi_1=0, \phi_2=2\pi/3, \phi_3=4\pi/3$, see FIG.\ref{figure3qubit}
We will prove that the optimal fidelity is upper bounded by $F_p$,
thus exactly equal to it.

By phase-covariant, we have the constraints,
\begin{eqnarray}
F_A(0)=F_A(2\pi/3)=F_A(4\pi/3)\nonumber
\end{eqnarray}
which mean,
\begin{eqnarray}
&&\lambda_1\cos(\psi_1-\frac{2\pi}{3})+\lambda_2\cos(\psi_2+\frac{2\pi}{3})+\lambda_3\nonumber\\
=&&\lambda_1\cos(\psi_1+\frac{2\pi}{3})+\lambda_2\cos(\psi_2-\frac{2\pi}{3})+\lambda_3\nonumber\\
=&&\lambda_1\cos(\psi_1)+\lambda_2\cos(\psi_2)+\lambda_3\nonumber .
\end{eqnarray}
Further they could be simplified as:
\begin{eqnarray}
&&\lambda_1\sin\psi_1=\lambda_2\sin\psi_2\nonumber\\
&&\lambda_1\cos\psi_1+\lambda_2\cos\psi_2=0.\label{lambda}
\end{eqnarray}
In symmetric cloning case, we assume this cloning machine works in the symmetric subspace.
So that $b=c$, $f=g$, and,
\begin{eqnarray}
F_A(\phi)=F_B(\phi)\equiv F.
\end{eqnarray}
This assumption is used in studying both PQCM and UQCM \cite{BCDM,BDEF}.
Therefore, fidelity for the three states can be written as,
\begin{equation}
F=\lambda_3=\frac{1}{2}+\frac{1}{2}(af^*+bh^*)
\end{equation}
The constraints may be simplified as follows,
\begin{eqnarray}
&&ab^*+ef^*+bd^*+fh^*=eb^*+fd^*,\nonumber\\
&&\mathrm{arg}(ab^*+ef^*+bd^*+fh^*)+\mathrm{arg}(eb^*+fd^*)=\pi,\nonumber\\
&&|a|^2+2|b|^2+|d|^2=1,\nonumber\\
&&|e|^2+2|f|^2+|h|^2=1,\nonumber\\
&&ae^*+2bf^*+dh^*=0.\label{Cons1}
\end{eqnarray}
We are seeking the optimal cloning machine, so we have to find the maximal fidelity.
By using algebraic inequalities, we find the fact,
$F\leqslant1/2+\sqrt{2}/{4}$,
see \cite{explain1} for detail. The equality
holds only when the following equations are satisfied,
$\mathrm{arg}(a)=\mathrm{arg}(f)$,
$\mathrm{arg}(b)=\mathrm{arg}(h)$,
$|a|=\sqrt{2}|f|$,
$|h|=\sqrt{2}|b|$,
$|e|=0$,
$|d|=0$.
From (\ref{Cons1}), we find
$2|b|^2+2|f|^2=1$,
$|b||f|=0$.
This implies, $\lambda_1=\lambda_2=0$.
And therefore,
\begin{equation}
F=\frac{1}{2}+\frac{\sqrt{2}}{4}.
\end{equation}
So we find that
the optimal cloning fidelity of a set of three equatorial qubits with equal
relative phases is exactly the optimal fidelity $F_p$ of the phase-covariant case.
All the possible parameters derived here are,
$|a|=1,|f|=\frac{1}{\sqrt{2}}$,$\mathrm{arg}(a)=\mathrm{arg}(f)$,$\mathrm{others}=0$, or $|h|=1,|b|=\frac{1}{\sqrt{2}}$,$\mathrm{arg}(b)=\mathrm{arg}(h)$,$\mathrm{others}=0$.
This is exactly the PQCM presented in \cite{BCDM}.  For completeness, we present it here explicitly,
\begin{eqnarray}
&&U|00\rangle=|00\rangle ,\nonumber\\
&&U|10\rangle=\frac {1}{\sqrt {2}}(|01\rangle+|10\rangle .
\label{pcloning}
\end{eqnarray}
We remark that this is the optimal cloning machine for only three equatorial qubits.

\begin{figure}[!h]
\includegraphics[width=9.3cm]{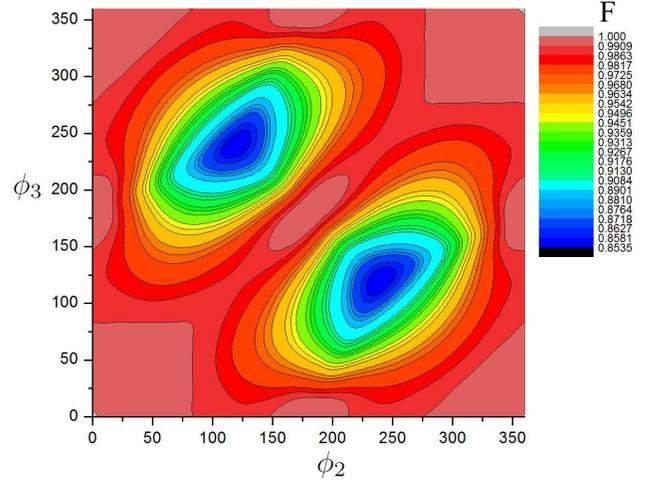}
\caption{(color online). Contour plot for fidelity of different $\phi _2$ and
$\phi _3$. Clearly the minimum points are ($\phi _2 = 120^\circ $,
$\phi _3 = 240^\circ $) and ($\phi _2 = 240^\circ $,
$\phi _3 = 120^\circ $), those two cases are equivalent.
}
\label{graph}
\end{figure}

Other three qubits on equator with different intersection angles do not have this characteristic. Figure.\ref{graph} shows some numerical result for different intersection angles. We set $\phi_1=0^\circ$, and the ranges of $\phi_2$ and $\phi_3$ are from $0^\circ$ to $360^\circ$.
We find unless $\phi_2=120^\circ, \phi_3=240^\circ$, the fidelity is always larger than $F_p$.
This is consistent with our analytic result. The numerical calculations are under
the conditions of equal fidelity and symmetric cloning.

Since we can clone arbitrary two equatorial qubits better than a PQCM does, we then
find the minimal input set determining completely a PQCM. Here we remark that
this is for economic case. Next we will show that ancillary states does not help to
increase the fidelity.

In order to solidify the equivalence between optimal three-state cloning machine
and the PQCM, we should prove it for the more general case
where the ancillary states are available since it is possible that we can
clone them better. The scheme of the analysis is almost
the same as that in economic case. Here we present the result and the detailed
analysis in the supplementary material. The cloning machine takes the form,
\begin{align}
U|00R\rangle &=a|00\rangle |0\rangle +b(|01\rangle+|10\rangle )|1\rangle \nonumber \\
U|10R\rangle &=f(|01\rangle +|10\rangle )|0\rangle +h|11\rangle |1\rangle ,
\end{align}
where  $|a|=\sqrt{2}|f|, |h|=\sqrt{2}|b|$.
A special case, $a=h=1/\sqrt{2}, b=f=1/2$, is identical to the PQCM in \cite{FIMW}.
The minimal set is also valid for the phase cloning of one to many case, $1\rightarrow n$.
When $n$ takes the limit of infinity, it corresponds to phase estimation \cite{phase-est}.
This implies that the uncertainty to find the value of an arbitrary phase
is equivalent as to identify a state in this minimal input set.
The detail of $1\rightarrow n$ phase cloning is presented in supplementary material.

\emph{Equivalence between 4-states cloning and a UQCM.}---A UQCM can copy
optimally an arbitrary qubit. It is known that we cannot do better than
a UQCM in cloning six states used in quantum cryptography \cite{Bruss}.
The problem is that whether the number of six states can be reduced to
the minimal sets which contains only five or even four states.
Considering that only three states can determine a PQCM, then it might
be possible that a UQCM, which has a lower fidelity than that of the PQCM,
may be determined by four input states. And this case must be the minimal input set.
We will show next that this is true.

\begin{figure}[!h]
\includegraphics[width=4.5cm]{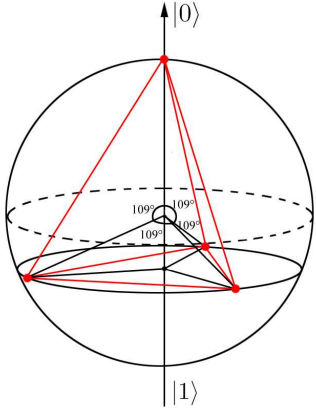}
\caption{(color online). Four states located on vertices of a inscribed tetrahedron
in the Bloch sphere can determine a UCQM.}
\label{figure4qubit}
\end{figure}

Let us consider four-states on the Bloch sphere with identical angular distance:
\begin{eqnarray}
|\psi_0\rangle&&=|0\rangle ,\nonumber\\
|\psi_1\rangle&&=\cos\frac{\theta}{2}|0\rangle+\sin\frac{\theta}{2}|1\rangle ,\nonumber\\
|\psi_2\rangle&&=\cos\frac{\theta}{2}|0\rangle+\sin\frac{\theta}{2}e^{i\frac{2\pi}{3}}|1\rangle ,\nonumber\\
|\psi_3\rangle&&=\cos\frac{\theta}{2}|0\rangle+\sin\frac{\theta}{2}e^{i\frac{-2\pi}{3}}|1\rangle ,
\label{four-state}
\end{eqnarray}
where $\theta$ satisfies $\cos\frac{\theta}{2}=\frac{\sqrt3}{3}$. And these four states form a tetrahedron,
see Figure (\ref{figure4qubit}).
We need to show that the optimal fidelity in cloning those four states is the same as that of a UQCM.

The general cloning machine can be assumed as,
\begin{align}
U|00R\rangle &=a|00A\rangle+b|01B\rangle+c|10C\rangle+d|11D\rangle \nonumber \\
U|10R\rangle &=e|00E\rangle+f|01F\rangle+g|10G\rangle+h|11H\rangle.
\label{g-cloning}
\end{align}
We could write down the fidelity $F(\phi)$ for $|\psi_1\rangle, |\psi_2\rangle, |\psi_3\rangle$:
\begin{equation}
F=\lambda_1\cos(2\phi+\psi_1)+\lambda_2\cos(\phi+\psi_2)+\lambda_3\nonumber
\end{equation}
where
$\lambda_1=2\sin^2\frac{\theta}{2}\cos^2\frac{\theta}{2}|ec^*\langle C|E\rangle+fd^*\langle D|F\rangle|$,
$\lambda_2=2\sin\frac{\theta}{2}\cos^3\frac{\theta}{2}|ea^*\langle A|E\rangle+fb^*\langle B|F\rangle+ac^*\langle C|A\rangle +bd^*\langle D|B\rangle|
+2\sin^3\frac{\theta}{2}\cos\frac{\theta}{2}|gc^*\langle C|G\rangle+hd^*\langle D|H\rangle+eg^*\langle G|E\rangle+fh^*\langle H|F\rangle| $,
$\lambda_3=(|a|^2+|b|^2)\cos^4\frac{\theta}{2}+(|g|^2+|h|^2)\sin^4\frac{\theta}{2}+(|e|^2+|f|^2+|c|^2+|d|^2)\sin^2\frac{\theta}{2}\cos^2\frac{\theta}{2}
+2\sin^2\frac{\theta}{2}\cos^2\frac{\theta}{2}|ag^*\langle G|A\rangle+bh^*\langle H|B\rangle|$.
Similar as in phase cloning case, requiring three states having equal fidelity would lead to
$F=\lambda_3$, see Eq. (\ref{lambda}).
The fidelity for input state $|\psi_0\rangle$ is obviously $F=|a|^2+|b|^2=\lambda _3$.

As usual, we assume the cloning machine work in symmetric subspace, that is, $b=c$, $f=g$, $|B\rangle=|C\rangle$, $|F\rangle=|G\rangle$.
After simplifying the expression of $F$, and considering $\cos\frac{\theta}{2}=\frac{\sqrt3}{3}$, we obtain,
\begin{eqnarray}
F=&&\frac{4}{9}-\frac{1}{9}(|a|^2+|b|^2)
+\frac{2}{9}(|f|^2+|h|^2)
\nonumber\\
&&
+\frac{4}{9}|af^*\langle F|A\rangle+bh^*\langle H|B\rangle|.
\label{g-fidelity}
\end{eqnarray}
By tricky but straightforward calculation, see supplementary, we can show that the fidelity $F$ is upper bound by, $F=5/6$,
which is exactly the optimal fidelity of a UQCM. This optimal fidelity is achievable, so we conclude that
the minimal input set of a UQCM contains only four states as presented in (\ref{four-state}) which
are located on vertices of a tetrahedron in FIG.\ref{figure4qubit}.

\emph{Conclusion.}---In summary, we have proved that the optimal cloner for three states with equal angular distances on the equator of the Bloch sphere is equivalent to the PQCM. This minimal set is also valid in the $1\rightarrow n$ cloning case.
For the UQCM, the minimal input set contains only four states located on vertices of a tetrahedron.
Those results sharpen further, and are important supplementaries to, the well-known results that the optimal quantum cloning machines
for BB84 states and six-state in quantum key distributions are PQCM and UQCM, respectively.

Since no-cloning and quantum cloning are fundamentals in quantum mechanics and quantum information,
it then seems natural to ask whether we can use those results for some quantum information applications, such as designing
some QKD protocols or quantum gambling. Also quantum cloning is related with
state-estimation or phase-estimation, we know that
our results actually provide the sets which have the highest uncertainty levels.
This may shed light on the study of uncertainty relationships in quantum mechanics.
One experimental application of those results is that, to test whether the cloning machines work,
we only need to check that those minimal input sets are cloned optimally. This will reduce dramatically the
testing procedure of the cloning machines.

\emph{Acknowledgements:} This work is supported by NSFC (10974247,11175248), ``973'' program (2010CB922904) and NFFTBS (J1030310).

\newpage

\newpage
\emph{Phase-covariant quantum cloning machine with ancillary states.}---Here we will show
that the ancillary state does not help to increase the fidelity in cloning three equatorial
qubits with equal relative phases.
A non-economic cloning machine is a unitary matrix acting on a larger Hilbert space with ancilla:
\begin{align}
U|00R\rangle &=a|00A\rangle+b|01B\rangle+c|10C\rangle+d|11D\rangle \nonumber \\
U|10R\rangle &=e|00E\rangle+f|01F\rangle+g|10G\rangle+h|11H\rangle.
\label{anci-cloning}
\end{align}
Similarly, we should have the orthogonal condition and normalization restrictions.
With assumption of symmetric space for quantum cloning,
we have $b=c$, $f=g$, $|B\rangle=|C\rangle$, $|F\rangle=|G\rangle$.
And we consider that the fidelity is invariant for different input qubits: $F(0)=F(2\pi/3)=F(4\pi/3)$.
The resulted fidelity has a similar form:
\begin{eqnarray}
F_A&=&\lambda_1\cos(2\phi+\psi_1)+\lambda_2\cos(\phi+\psi_2)+\lambda_3
\nonumber \\
&=&\frac{1}{2}+\frac{1}{2}Re(af^*\langle F|A\rangle+bh^*\langle H|B\rangle),
\end{eqnarray}
where we use the notations,
$\lambda_1=\frac{1}{2}|ec^*\langle C|E\rangle+fd^*\langle D|F\rangle|$,
$\psi_1=\arg(ec^*\langle C|E\rangle+fd^*\langle D|F\rangle)$,
$\lambda_2=\frac{1}{2}|ac^*\langle C|A\rangle+eg^*\langle G|E\rangle+bd^*\langle D|B\rangle+fh^*\langle H|F\rangle|$,
$\psi_2=\arg(ac^*\langle C|A\rangle+eg^*\langle G|E\rangle+bd^*\langle D|B\rangle+fh^*\langle H|F\rangle)$,
$\lambda_3=\frac{1}{2}Re(ag^*\langle G|A\rangle+bh^*\langle H|B\rangle)+\frac{1}{2}$.
Consider the restrictions mentioned above,
similar as the economic cloning case, we find the the maximal fidelity is,
$F=\frac{1}{2}+\frac{1}{\sqrt{8}}$, which is obtained at $|a|=\sqrt{2}|f|, |h|=\sqrt{2}|b|, |d|=|e|=0$.
Due to the presence of ancillary states. The restrictions can be rewritten as,
\begin{eqnarray}
&&2|b|^2+2|f|^2=1\nonumber\\
&&ab^*\langle B|A\rangle+fh^*\langle H|F\rangle=0\nonumber\\
&&\mathrm{arg}(ab^*\langle B|A\rangle+fh^*\langle H|F\rangle)=\pi\nonumber\\
&&2bf^*\langle F|B\rangle=0.\label{Cons3}
\end{eqnarray}
If we set $|A\rangle=|B\rangle=|F\rangle=|H\rangle$, then the cloning machine reduces to the economic case.
But if we set $|A\rangle=|F\rangle=|0\rangle$, $|B\rangle=|H\rangle=|1\rangle$, then $\langle B|A\rangle=\langle H|F\rangle=0$,
the only restriction is,
\begin{eqnarray}
2|b|^2+2|f|^2=1.
\end{eqnarray}
Under this condition, the non-economic quantum cloning is always optimal. Explicitly they can
take the following form (\ref{anci-cloning}),
\begin{align}
U|00R\rangle &=a|00\rangle |0\rangle +b(|01\rangle+|10\rangle )|1\rangle \nonumber \\
U|10R\rangle &=f(|01\rangle +|10\rangle )|0\rangle +h|11\rangle |1\rangle .
\end{align}
Note that $|a|=\sqrt{2}|f|, |h|=\sqrt{2}|b|$. This form is more general than
the well-known PQCM,
a special case, $a=h=1/\sqrt{2}, b=f=1/2$, is identical to the PQCM in \cite{FIMW}.

\emph{The $1\rightarrow n$ PQCM.}---Next similar as in case of cloning one state to
two copies, we will show that the  $1\rightarrow n$ PQCM, which can clone one state to $n$ copies,
can be determined by these three equatorial qubits presented in our main text.

For $1\rightarrow n$ case, we still assume that our cloning machine is working in symmetric subspace,
and it is economic. The transformations can be expressed as,
\begin{eqnarray}
|0\rangle\longrightarrow \sum^n_{i=0}a_i|i\rangle\rangle ,\nonumber\\
|1\rangle\longrightarrow \sum^n_{i=0}b_i|i\rangle\rangle ,
\end{eqnarray}
where $|i\rangle\rangle$ is a complete symmetric state with $i$ states in $|1\rangle$ among all $n$ qubits.
For example, if $n=3$, $|1\rangle \rangle \equiv (|001\rangle +|010\rangle +|100\rangle )/\sqrt {3}$. 
Analogously as in $1\rightarrow 2$ case, parameters should satisfy the constraints,
$\sum_{i=0}^n|a_i|^2=1$, $\sum_{i=0}^n|b_i|^2=1$, and $\sum_{i=0}^na_i b_i^*=0$.

The input is a equatorial qubit, $|\psi\rangle =\frac{1}{\sqrt{2}}(|0\rangle+\mathrm{e}^{\mathrm{i}\phi}|1\rangle)$,
we find the output state by cloning transformations, $\sum^n_{i=0}(a_i+\mathrm{e}^{\mathrm{i}\phi}b_i)|i\rangle\rangle$.
Without loss of generality, tracing off all states except the first one,
we can obtain the one-qubit reduced density matrix, $\rho _1$.
By complicated but straightforward calculations, the fidelity can be found as,
\begin{eqnarray}
F&=&
\frac{1}{2}+\frac{1}{2}Re(\sum^{n-1}_{i=0}(a_ia_{i+1}^*\mathrm{e}^{\mathrm{i}\phi}+b_ib_{i+1}^*\mathrm{e}^{\mathrm{i}\phi}\nonumber\\
&&+a_ib_{i+1}^*+a_{i+1}^*b_i\mathrm{e}^{2\mathrm{i}\phi}))\frac{\sqrt{(n-i)(i+1)}}{n}.
\end{eqnarray}
Here, we have used the following identities to simplify our expression,
$C_{n-1}^i+C_{n-1}^{i+1}=C_n^i$, $\frac{C_{n-1}^i}{\sqrt{C_n^iC_n^{i+1}}}=\frac{\sqrt{(n-i)(i+1)}}{n}$.
Therefore, as in the 1$\rightarrow$2 case, we express the fidelity as,
\begin{equation}
F=\lambda_1\cos(2\phi+\psi_2)+\lambda_2\cos(\phi+\psi_1)+\lambda_3,
\end{equation}
where $\lambda_1=\frac{1}{2}\sum_{i=0}^{n-1}|a_{i+1}^*b_i|$,
$\lambda_2=\frac{1}{2}\sum_{i=0}^{n-1}|a_ia_{i+1}^*+b_ib_{i+1}^*|$ and
$\lambda_3=\frac{1}{2}\sum_{i=0}^{n-1}|a_ib_{i+1}^*|+\frac{1}{2}$.
Similarly, when three states are cloned equally well, we have
$\lambda_1=\lambda_2$,
$\psi_1+\psi_2=\pi,(k\in\mathbb{Z})$.
So that fidelity for them is $F=\lambda_3$

Next, we look for the maximal fidelity for them and find the parameters,
\begin{eqnarray}
F&=&\frac{1}{2}+\frac{1}{2}Re(\sum^{n-1}_{i=0}a_ib_{i+1}^*\frac{\sqrt{(n-i)(i+1)}}{n})\nonumber\\
&\leq&\frac{1}{2}+\frac{1}{4}\sum^{n-1}_{i=0}(|a_i|^2+|b_{i+1}|^2)\frac{\sqrt{(n-i)(i+1)}}{n}.\nonumber
\end{eqnarray}
By considering the normalization conditions for $a_i$ and $b_i$,
we have the following results,
\begin{eqnarray}
F&\leq &\frac{1}{2}+\frac{\sqrt{n(n+2)}}{4n}, ~~~~{\rm n~is~even;}\\
F&\leq &\frac{1}{2}+\frac{n+1}{4n}, ~~~~~~~~~~~~~~{\rm n~is~odd;}
\end{eqnarray}
For $n$ is even, ``='' is satisfied only when the following equations are satisfied,
$\mathrm{arg}(a_i)=\mathrm{arg}(b_{i+1})$,
$|a_{\frac{n}{2}}|=|b_{\frac{n}{2}+1}|=1$, other parameters are zeroes;
For $n$ is odd, ``='' is satisfied only when the following equations are satisfied,
$\mathrm{arg}(a_i)=\mathrm{arg}(b_{i+1})$, $|a_{\frac{n-1}{2}}|=|b_{\frac{n+1}{2}}|=1$,
other parameters are zeroes. Note that now $\lambda_1=\lambda_2=0$.
Those results agree with the results for $1\rightarrow n$ phase-covariant cloning machine in \cite{1-Nphase}.
So we conclude that three equatorial qubits with equal relative phases can determine
completely the optimal $1\rightarrow n$ PQCM.
Obviously, this general result is consistent with our previous $n=2$ case.
Our result is also true in case, $n\rightarrow \infty$.
The implication of this result is that, as stated in our main text, to identify one state from
the minimal set which contains three states is as difficult as to
find the exact
value of the phase in a equatorial qubit. This is quite surprising.

\emph{Further calculation of equation (17) and optimal UQCM.}---From Eq.(17), we have,
\begin{eqnarray}
F+\frac{1}{3}F\leq &&\frac{4}{9}-\frac{1}{9}(|a|^2+|b|^2)+\frac{2}{9}(|f|^2+|h|^2)\nonumber\\
&&+\frac{4}{9}(|a||f|+|b||h|)+\frac{1}{3}(|a|^2+|b|^2)\nonumber\\
\leq &&\frac{4}{9}+\frac{2}{9}(|a|^2+|b|^2+|f|^2+|h|^2)\nonumber\\
&&+\frac{4}{9}(\frac{|a|^2+4|f|^2}{4}+\frac{4|b|^2+|h|^2}{4})\nonumber\\
=&&\frac{10}{9}.
\end{eqnarray}
So we have, $F\le \frac{5}{6}$, as shown in the main text.
Then maximal fidelity equals to the fidelity of 2 dimensional UQCM.
The equality "=" is satisfied only when $|a|=|h|=\sqrt{\frac{2}{3}}$, $|b|=|f|=\sqrt{\frac{1}{6}}$, $|d|=|e|=0$. Consider the constraints, we have $\langle B|F\rangle=0$. Hence we got the only possible form of $|A\rangle, |B\rangle, |F\rangle, |H\rangle$: $|A\rangle=|0\rangle, |B\rangle=|1\rangle, |F\rangle=|0\rangle, |H\rangle=|1\rangle$ with some possible phase factors (The requirement is $Im(af^*\langle F|A\rangle)=Im(bh^*\langle H|B\rangle)=0$).
This is indeed the well-known UQCM.
Hence we proved optimal cloning machine of 4 states which form a tetrahedron is equivalent to UQCM.


\begin{thebibliography}{99}

\bibitem{nocloning} W. K. Wootters and W. H. ZureK, Nature (London) {\bf299}, 802(1982).

\bibitem{Nielsen-Chuang2000} M. A. Nielsen and I. C. Chuang, {\it Quantum computation
and quantum information}, Cambridge University Press, Cambridge
2000.

\bibitem{UQCM} V. Bu\v{z}ek and M. Hillery, Phys. Rev. A {\bf54}, 1844 (1996).

\bibitem{Werner} R. F. Werner, Phy. Rev. A {\bf58}, 1827 (1998).


\bibitem{d_dim}V. Bu\v{z}ek and M. Hillery, Phys. Rev. Lett. {\bf81}, 5003 (1998).

\bibitem{BDEF}D. Bru\ss ,
D. DiVincenzo, A. Ekert, C. A. Fuchs, C. Macchiavello, and J. A.
Smolin,
Phys. Rev. A {\bf 57}, 2368 (1998).

\bibitem{GM1997} N. Gisin and S. Massar, Phys. Rev. Lett. {\bf 79},
2153 (1997).

\bibitem{review}N. Gisin, G. Ribordy, W. Tittel, and H. Zbinden, Rev. Mod. Phys. {\bf 74}, 145 (2002).

\bibitem{CerfSecurity}N. J. Cerf, M. Bourennane, A. Karlsson, and N. Gisin, Phys. Rev. Lett. {\bf 88}, 127902(2002).


\bibitem{Fan}H. Fan, K. Matsumoto, and M. Wadati, Phys. Rev. A {\bf64}, 064301 (2001).


\bibitem{BCDM}D. Bru\ss , M. Cinchetti, G. M. D'Ariano, and C. Macchiavello, Phys. Rev. A {\bf 62}, 012302 (2000).

\bibitem{TwoBases}D. Bru\ss  ~and C. Macchiavello, J. Phys. A {\bf34}, 6815 (2001).

\bibitem{FGGNP}C. A. Fuchs, N. Gisin, R. B. Griffiths, C.-S. Niu, and A. Peres, Phys. Rev. A {\bf56}, 1163 (1997).

\bibitem{1-Nphase}H. Fan, K. Matsumoto, X. B. Wang, and M. Wadati, Phys. Rev. A {\bf65} 012304 (2001).

\bibitem{FIMW}H. Fan, H. Imai, K. Matsumoto, and X. B. Wang, Phys. Rev. A {\bf67}, 022317(2003).

\bibitem{Bechmann}H. Bechmann-Pasquinucci and N. Gisin, Phys. Rev. A {\bf59}, 4238(1998).

\bibitem{Wangunified}Y. N. Wang, H. D. Shi, Z. X. Xiong, L. Jing, X. J. Ren, L. Z. Mu, and H. Fan, Phys. Rev. A {\bf 84}, 034302 (2011).

\bibitem{Xiong}Z. X. Xiong, H. D. Shi, Y. N. Wang, L. Jing, J. Lei, L. Z. Mu, and H. Fan, Phys. Rev. A {\bf85}, 012334 (2012).


\bibitem{BB84}C. H. Bennett and G. Brassard, in Proceedings of the IEEE International Conference on Computers, Systems and Signal Processing, Bangalore, India (IEEE, New York, 1984), pp. 175-179.







\bibitem{Bruss}D. Bru\ss, Phys. Rev. Lett. {\bf 81}, 3018(1998).


\bibitem{naturephotonics09}Nagali, E. \emph{et al}.
Nature Photonics {\bf3}, 720 (2009).

\bibitem{Recent1}G.M.D'Ariano, S. Facchini, and P. Peritotti, Phys. Rev. Lett. {\bf 106}, 010501 (2011).
\bibitem{Recent2}P. Kurzynski  \emph{et al}., Phys. Rev. Lett. {\bf 106}, 180402 (2011).
\bibitem{Recent3}G. Qin  \emph{et al}., Phys. Rev. Lett. {\bf 106}, 180404 (2011).
\bibitem{Recent4}J. Bae, W. Y. Hwang, and Y. D. Han, Phys. Rev. Lett. {\bf 107}, 170403 (2011).
\bibitem{Recent5}S. Raeisi, W. Tittel, and C. Simon, Phys. Rev. Lett. {\bf 108}, 120404 (2012).
\bibitem{Recent6}M. M. Wilde, P. Hayden, and S. Guha, Phys. Rev. Lett. {\bf 108}, 140501 (2012).
\bibitem{Recent7}G. Smith and J. A. Smolin, Phys. Rev. Lett. {\bf 108}, 230507 (2012).
\bibitem{state}M. Paris, J \v{R}eh\'{a}\v{c}ek (Eds.), Quantum State Estimation, Lect. Notes Phys. {\bf649} (Springer, Berlin Heidelberg 2004).

\bibitem{phase-est}W. van Dam, G. M. D'Ariano, A. Ekert, C. Macchiavello,
and M. Mosca, Phys. Rev. Lett. {\bf 98}, 090501 (2007).

\bibitem{explain1}\begin{eqnarray}
F&=&\frac{1}{2}+\frac{1}{2}(af^*+bh^*),\nonumber\\
&=&\frac{1}{2}+\frac{1}{2}(|a||f|\cos(\mathrm{arg}(a)-\mathrm{arg}(f))\nonumber\\
&&+|b||h|\cos(\mathrm{arg}(b)-\mathrm{arg}(h)))\nonumber\\
&\le &\frac{1}{2}+\frac{1}{2}(|a||f|+|b||h|)\nonumber\\
&\le &\frac{1}{2}+\frac{1}{4\sqrt{2}}(|a|^2+2|f|^2+|h|^2+2|b|^2)\nonumber\\
&\le &\frac{1}{2}+\frac{\sqrt{2}}{4}\nonumber
\end{eqnarray}

\end{thebibliography}
\end{document}